\documentclass[dvips,12pt]{article}
\begin{document}

\title{Motion of Spin $1/2$ Massless Particle in
a Curved Spacetime.\\ I. Lagrangian Approach}
\date{}

\author{\footnotesize A.T. Muminov\\\footnotesize
Ulug-Beg Astronomy Institute, Astronomicheskaya~33,
Tashkent 100052, Uzbekistan\\
\footnotesize amuminov@astrin.uzsci.net}

\maketitle

\begin{abstract}
Quasi-classical picture of motion of spin $1/2$ massless particle
in a curved spacetime is built on base of simple Lagrangian model.
The one is constructed due to analogy with Lagrangian of massive
spin $1/2$ particle \cite{al}. Equations of motion and spin
propagation coincide with Papapetrou equations describing dynamic
of photon in a curved spacetime~\cite{za}
\end{abstract}

%================ definitions
%
\def \oplus {\lefteqn{\hspace{0.053em}\bigcirc}{\,+}{}\hspace{0.5em}}
\def \Realset {\mbox{R}}
\def \vc {\vec}
\newcommand{\T}{\tilde }
\newcommand{\comment}[1]{}
\newcommand{\lrr}[1]{\left(#1\right)}
\newcommand{\lrc}[1]{\left\{#1\right\}}
%\newcommand{\rr}{\right)}
%\newcommand{\lfc}{\left\{}
%\newcommand{\rtc}{\right\}}
%\newcommand{\lta}{\left\langle}
%\newcommand{\rta}{\right\rangle}
%!!!!!!!!!!!!!!!!!!!!!!!!!To REMOVE or REDEFINE!!!!!!!!!!
%\def \om {\omega}
\def \wdg {\wedge}
\def \vcx {\dot{\vc x }}
\def \dtx {\dot{x}}
\def \dcn {\mbox{D.C. }}
\newcommand{\lcom}[1]{{\left[#1\right.}}
\newcommand{\rcom}[1]{{\left.#1\right]}}
\newcommand{\lacm}[1]{{\left\{#1\right.}}
\newcommand{\racm}[1]{{\left.#1\right\}}}
\newcommand{\cc}[1]{\bar{#1}}

\newcommand{\ihlf}{\frac{i\hbar}{2}}
\newcommand{\ihqr}{\frac{i\hbar}{4}}
\def \qrt {{1\over4}}
\def \hspin {{1\over2}}
\def \hsp {1/2\,}
%!!!!!!!!!!!!!!!!!!!!!!!!!!!!!!!!!!!!!!!!!!!!!!!!!!!!!!!!!!!!
%\newcommand{\too}{\longrightarrow}
\newcommand{\eps}{\varepsilon}
\newcommand{\fracm}[2]
{{\displaystyle#1\over\displaystyle#2\vphantom{{#2}^2}}}
\newcommand{\prt}{\partial\,}
\def \Lge {{\cal L}}
\def \vn {{\vec{n}}}
\def \cdt {\hspace{0.1em}\cdot\hspace{0.1em}}
\def \fdt {\hspace{0.4em}}
\newcommand{\con}[2]{\omega_{#1\cdt}^{\fdt#2}}
\newcommand{\cur}[2]{\Omega_{#1\cdt}^{\fdt#2}}
\newcommand{\dd}[2]{\frac{\prt\,#1}{\prt\,#2}}
\newcommand{\pp}[2]{{\prt\,#1}/{\prt#2}}
\newcommand{\dds}[1]{\frac{d\,#1}{ds}\,}
\newcommand{\Ds}[1]{\frac{D\,#1}{ds}\,}
%-----
%\def \vep {\vc{{}\,\eps}\,}
%\def \vn {\vc{{}\,n}}
\def \bps {{\psi^\dagger}}
\def \dps {\dot{\psi}}
\def \dbps {\dot{\bps\,{}}}
\def \gam {{\hat{\sigma\,{}}\!{}}}
\def \onh {$1/2$ }
\def \frm #1  {\{#1\}\,{}}
%========================================

\section{Introduction}

Lagrangian describing motion of non scalar particles in an
external field includes additional terms containing spin
variables. These terms change equation of motion of the particle
due to interaction of spin with the external field \cite{frnk}.
Motion of particles with spin in curved spacetime was subject of
studies of numerous authors \cite{papa,Dix,zfr}, see more in
review \cite{frszk}. It was shown in the mentioned works that
equation of motion differs from geodesic equation due to a term
which is contraction the curvature with spin and velocity.
However, there is still no general consensus on the behavior of
particle with spin in external gravitation field. Certain progress
on the description of massive particles with spin $1$ and $1/2$
was achieved in our recent works \cite{zr1,al}. At the same time,
as it was pointed in our work \cite{za} description of massless
particles cannot be derived straightforwardly from the Lagrange
description of massive particles just by equating the mass to
zero: $m=0.$ Thus, development Lagrange description of motion of
spin $1/2$ massless particle in a curved spacetime needs especial
consideration which is given in the present work.

We consider the particle as quasi-classical. In other words this
means that motion of the particle is described by its coordinates
$\{x^i\}$ and internal spin variables specifying spin degrees of
freedom of the particle. Spin variables are elements of suitable
spinor spaces. Determination of the spaces demands specifying an
orthonormal frame in considered domain of spacetime.
\comment{ Since we consider a massless particle, we can not
introduce comoving frame along the worldline by the same way as
for massive particle \cite{al}.}
It is seen that procedures of constructing a field of comoving
frames along null-curve and timelike curve sufficiently differ
from each other. In case of massive particle we could use length
parameter $s$ along the worldline which specified timelike vector
of unit length $\vn_0=\dtx^i\prt_i=\vcx,$ $\dtx^i=dx^i/ds.$ The
vector defined $1+3$ splitting of tangent space along the
worldline used for specifying spinor variables \cite{al}. In case
of massless particle vector $\vn_-=\vcx$ tangent to worldline of
the particle has zero length and can not be normed by the same way
as timelike vector. We need second null-vector, say $\vn_+,$ which
will norm the vector $\vn_-$ by the condition $<\vn_-,\vn_+>=1.$
This determines $2+2$ splitting of tangent space along the
worldline. Let $\{ \vn_\alpha\},$ $\alpha=1,2$ two spacelike
vectors which supplements coframe $\frm \vn_\pm$ up to frame of
tangent space along the worldline.
It was shown in our work \cite{za} that vector $\vn_+$ specifies
frequency of oscillations $\omega$ of wave function of the
particle which defines energy of the particle $E=\hbar\omega$
referred to the frame. Vectors $\vn_\alpha$ specify polarization
subspace of tangent space. Both pairs of vectors $\vn_\pm$ and
$\vn_{1,2}$ are defined with accuracy up to arbitrary rotations
which are elements of groups $SO(1,1)$ and $SO(2)$ accordingly.

Rotating vectors $\vn_\pm$ such a way that value of energy $E$ in
the frame becomes a constant along the worldline as it was done
under consideration of electromagnetic field \cite{za}, we define
canonical velocity $\vcx=\vn_-$ and canonical parameter $s$ for
this null-worldline. Form of the Lagrangian and equations of
motion of the particle take their simplest form under choice of
the canonical parameter as time parameter along the worldline.
Procedure of variation the action $S=\int\Lge\,ds $ assumes
independent variation of $\vcx$ and spinor variables under
constraint $E=const$. Hence, spinor variables should be referred
to coframe $\frm \vn_\alpha $ of vectors of polarization. In other
words, spinor variables are elements of linear spaces of
representation of group $SO(2)$.

%-->
\comment{It is convenient to chose decomposable spaces of
representation of group $SO(2)$ which are (indecomposable) spaces
of representation of group $SO(3)\!\supset\!SO(2).$ Procedure of
construction the spaces was explained in work \cite{al}. In brief,
first of all Pauli matrices $\frm \sgm^1,\sgm^2,\sgm^3 $ referred
to frame $\frm \vn_1,\vn_2,\vn_3$ are introduced. Here
$\vn_3=2^{-1/2}(\vn_-+\vn_+).$}
%<---

The spaces are constructed as follows. Pauli matrices $\{
\gam^\alpha\},$ ${}\,\alpha=1,2$  referred to frame of
polarization are introduced. The matrices generate local Clifford
algebras referred to the frame. Besides, local Clifford algebra
introduced this way specifies two local spinor spaces attached to
the worldline. These spaces are two spaces of spinor
representations of the group $SO(2)$ which are isomorphic to each
other under Hermitian conjugation. In our approach elements of the
spaces $\bps,\psi$ play role of generalized coordinates which
describe spin degrees of freedom  of the particle.

The desired Lagrangian should depend on generalized coordinates
$\frm x^i ,$ $\frm {\bps,\psi} $ and their derivatives over $s$.
At the same time the Lagrangian should contain covariant
derivatives of spinor variables $\bps$ and $\psi$. Moreover, the
Lagrangian must contain term with $<\vcx,\vcx>$ which yields left
hand side (LHS) of geodesic equation. Euler-Lagrange equations for
spinor variables are expected to yield reduced form of Dirac
equation for wave propagating along the worldline of the particle.
In the limiting case of zero gravitation the equation becomes a
consequence of Weyl equation formulated for massless spinor field.
All these requirements determine the form of the Lagrangian
describing motion of massless spin \onh particle in curved
spacetime. Thus, Euler-Lagrange equations are reduced to equations
describing motion of the spin along the particle worldline and the
worldline shape. The equations obtained this way are similar to
Papapetrou equations for photon \cite{za}.
%\newpage

\section{Lagrange formalism for massless particle of spin $1/2$}

First of all, we have to specify a field of orthonormal frames
along the worldline. For this end we add one more null-vector
$\vn_+$ transversal to null-vector $\vn_- =\vcx$. The vectors
$\vn_\pm$ norm each other such a way that $<\vn_-,\vn_+>=1.$ This
defines $2+2$ splitting of tangent spaces
$T=(\Realset\vn_-\oplus\Realset\vn_+)
\oplus(\Realset\vn_1\oplus\Realset\vn_2)$ on the worldline.
Vectors $\frm \vn_a ,$ $a=\pm,1,2$ constitute comoving orthonormal
frame along the worldline. Spacelike vectors $\{\vn_\alpha\},$
$\alpha=1,2$ constitute frame of polarization subspace of tangent
space. The vector $\vn_+$ defines the value of energy of the
particle $E=\hbar\omega$ for frame in question \cite{za}. Vectors
$\vn_\pm$ are defined with accuracy of arbitrary  Lorentzian
rotations in subspace $\Realset\vn_-\oplus\Realset\vn_+.$ We fix
the vectors by condition $E=const$ along the worldline which, in
turn, defines canonical parameter along the worldline \cite{za}.
Let $\frm \nu^a $ be covector comoving frame dual to $\{\vn_a\}.$
It is seen that $\frm \nu^\alpha$ is covector polarization frame
dual to vector frame $\frm \vn_\alpha $ in tangent polarization
subspace.

%\newpage
%Then following \cite{al} we construct two spinor spaces of
%representation of group $Spin(3)$ and $SO(3)$.
In order to describe spin variables of the Lagrangian we introduce
Pauli matrices $\frm \gam^1,\gam^2 $ referred to the covector
polarization frame $\{\nu^\alpha\}$. The matrices are constant in
chosen frame and obey anticommutation rules as follows:
\begin{equation}\label{gam-rules}
\{\gam^\alpha,\gam^\beta\}=-2\eta^{\alpha\beta},
\end{equation} where $\eta^{\alpha\beta}$ is Minkowski tensor in
subspace $\Realset\vn_1\oplus\Realset\vn_2.$ Algebraic span of
Pauli matrices yields local sample of Clifford algebra in each
point of the worldline. Union of the local Clifford algebras
constitute fibre bundle of Clifford algebra along the worldline.

Invertible elements $R$ of Clifford algebra such that
$$ R^{-1}=R^\dagger,
$$where $R^\dagger$ stands for Hermitian conjugated matrix, constitute
$Spin(2)$ group \cite{ccl}. There is an endomorphism $R:SO(2)\to
Spin(2)$ defined by formula:
\begin{equation}\label{trans}
R_L\gam^\alpha R^{-1}_L=L_{\beta\cdt}^{\fdt\alpha}\gam^\beta,\quad
(L_{\beta\cdt}^{\fdt\alpha})\in SO(2).
\end{equation}
Elements of local Clifford algebra are operators on two local
spinor spaces referred to the frame on the worldline. The spaces
are local linear spaces of representation of group $Spin(2)$ and
$SO(2)$. Elements of the local spaces $\psi\in S$ and $\bps\in
S^\dagger$ are $2\times1$ and $1\times2$ complex matrices
accordingly.  This way  element $L$ of group of spatial rotations
$SO(2)$ acts on spaces of representation of the group as follows:
\begin{equation}\label{spstrans} '\psi=R_L\psi,\quad
'\bps=\bps R_L^{-1},\quad \psi\in S,\,\bps\!\in S^\dagger.
\end{equation}
Union of the local spinor spaces constitute spinor fibre bundle on
the worldline.

Image of an infinitesimal rotation $L={\bf1}-\eps\in SO(2)$ is:
\begin{equation}\label{RInf}
R_{1-\eps}=\hat{1}+\delta
Q=\hat{1}+1/4\,\eps_{\alpha\beta}\,\gam^\alpha \gam^\beta.
\end{equation} The infinitesimal transformation rotates elements of
the rest frame:
\begin{equation}\label{rot-cvec}
\delta\nu^\alpha=-\eps_{\beta\cdt}^{\fdt \alpha}\nu^\beta.
\end{equation}
Accordingly (\ref{spstrans}) the rotation initiates a
transformation of spinors:
\begin{equation}\label{spinRot}
\delta\psi=1/8\,\eps_{\beta\gamma}\,[\gam^\beta,\gam^\gamma]\,\psi,\quad
\delta\bps=-1/8\,\eps_{\beta\gamma}\,\bps\,[\gam^\beta,\gam^\gamma],
\end{equation}
under which  Pauli matrices rotates as follows:
\begin{eqnarray*}
'\gam^\alpha=R\gam^\alpha R^{-1},\quad \delta\gam^\alpha=[\delta Q,\gam^\alpha];\\
\delta\gam^\alpha=-\eps_{\beta\cdt}^{\fdt \alpha}\gam^\beta=
1/4\eps_{\beta\gamma}\left[\gam^\beta\gam^\gamma,\gam^\alpha\right].
\end{eqnarray*}
It is seen that the rotation coincides with rotation of components
of contravariant vector with accuracy up to opposite sign. Thus,
if we take into account both of the transformations Pauli matrices
become invariant as it is accepted in field theory \cite{Besse}.
%However, since we consider mechanical model with detailed
%direction of motion we will use only one of the transformations
%which is given by (\ref{gamRot}).

State of the particle is described by its coordinates $\frm x^i $
in spacetime, spinor variables $\frm {\psi,\bps} $ which are
elements of spinor fibre bundles on the worldline and their
derivatives $\dtx^i={dx^i\over ds}$ and $\dds{\psi}$, $\dds{\bps}$
over canonical parameter $s$ along the worldline. Polarization
frame rotates along the worldline:
$$\dot\nu^\alpha =-\con{\beta}{\alpha}(\vcx)\nu^\beta ,
$$
where angular velocities are given by values of Cartan' rotation
1-forms $\con{\beta}{\alpha}=\gamma_{c\beta\cdt}^{\fdt\fdt
\alpha}\nu^c$ on vector $\vcx$. So do spinor variables referred to
the frame. Their transformations are given by equations
(\ref{spinRot}) where
$\eps_{\beta\cdt}^{\fdt\alpha}=\gamma_{-\beta\cdt}^{\fdt\fdt\alpha}$.
The transformations are taken into account by covariant
derivatives of spinor variables along the worldline:
\begin{eqnarray}\label{ders}
\dps=\dds{\psi}+\qrt\gamma_{b\delta\eps}\,\dtx^b\,\gam^\delta\gam^\eps \psi,\\
\dbps=\dds{\bps}-\qrt\gamma_{b\delta\eps}\,\dtx^b\,\bps\,\gam^\delta\gam^\eps.
\nonumber
\end{eqnarray} Besides, total covariant derivatives
(with account of spinor transformation and rotation of vector
indexes) of Pauli matrices are zero.

Since Lagrangian of the particle is covariant under internal
transformations of the polarization frames, it includes
derivatives (\ref{ders}). The Lagrangian contains term
$E/2\,\bps\psi<\vcx,\vcx>$ which yields geodesic equation.
Euler-Lagrange equations for spin variables should lead to reduced
form of Dirac equation. Analysis shows that to satisfy such the
requirement we may accept the following form of the Lagrangian:
\begin{equation}\label{lagr}
\Lge={E\over2}\,\bps\psi<\vcx,\vcx>-\ihlf\left(\bps\dps-\dbps\psi\right)
.
\end{equation}
% E-energy
% MORE COMMENT !!!!!!!!!!!!!!!!!
%
We must bear in mind that the Lagrangian is function of
generalized coordinates $x^i,$ $\psi,$ $\bps$ and their velocities
$\dds{x^i}=\dtx^i,$ $\dds{\psi},$ $\dds{\bps}$. At the same time
covariant form of the Lagrangian includes derivatives represented
in orthonormal frame. Due to this we recall formulae of
transformations between the frames:
\begin{equation}\label{frtrans}
\vn_a=n_a^i\pp{\!}{x^i},\quad \dtx^a=n^a_i\dtx^i,
\quad\dtx^i=n^i_a\dtx^a,\quad n^i_an_j^a=\delta^i_j, \quad
n^i_an^b_i=\delta_a^b.
\end{equation}

\section{Euler-Lagrange equations for spinor variables}

Due to (\ref{lagr}) generalized momenta  conjugated to generalized
coordinates $\bps$ and $\psi$ are:
$$
\Psi=\pp{\Lge}{\!{}\!\lrr{\dds{\bps}}}=\ihlf\, \psi,\quad
\Psi^\dagger=\pp{\Lge}{\!{}\!\lrr{\dds{\psi}}}=-\ihlf\,\bps .
$$
Euler Lagrange equations for the considered generalized
coordinates read:
\begin{equation}\label{ELE1}
\dds{}\Psi=\pp{\Lge}{\bps},\quad
\dds{}\Psi^\dagger=\pp{\Lge}{\psi}.
\end{equation}
Straightforward calculation of the right hand side (RHS) of the
above equations gives:
\begin{eqnarray*}
\pp{\Lge}{\bps}=-\ihlf
\dps-\ihlf\cdot{1\over4}\gamma_{b\delta\eps}
\dtx^b\,\gam^\delta\gam^\eps\psi,\\
\pp{\Lge}{\psi}=\ihlf\dbps
-\ihlf\cdot{1\over4}\gamma_{b\delta\eps}
\dtx^b\,\bps\gam^\delta\gam^\eps.
\end{eqnarray*}
Now it is seen that the  RHS of the equations (\ref{ELE1})
completes ordinary derivatives of spinor variables in the LHS up
to covariant derivatives. This way Euler-Lagrange equations for
$\bps,\psi$ generalized coordinates become:
\begin{equation}\label{eq4psi}
i\hbar \dps=0, \quad{}\quad i\hbar\dbps =0.
\end{equation}
The equations coincide with Weyl \comment{??} equations for
massless freely propagating spinor field of definite helicity
\cite{NN}.

\section{Generalized momentum conjugated with $x^i$ and
conservation of spin}

Generalized momentum conjugated with $x^i$ is
$p_i=\pp{\Lge}{\dtx^i}.$ However it is convenient to operate with
generalized momenta expressed in orthonormal frame:
$p_a=n_a^ip_i.$ Differentiating (\ref{lagr}) over $\dtx^a$ we
obtain:
\begin{eqnarray*}p_a=\dd{\Lge}{\dtx^a}=E\bps\psi
\,\eta_{ab}\dtx^b-\ihlf\cdot{1\over2}\gamma_{a\delta\eps}\,\bps
\gam^\delta\gam^\eps\psi.
\end{eqnarray*}
We define spin of the particle as:
\begin{equation}\label{spin}
S^{\delta\eps}=-\ihqr\,\bps\gam^\lcom{\delta}\gam^\rcom{\eps}\psi,
\end{equation}
where $[,]$ stands for commutator of the Pauli matrices. By
construction, the spin is element of space which is tensor product
of two copies of tangent polarization space. Thus, it has no
$\pm$-components and condition of orthogonality of the spin to
velocity is satisfied: $\dtx^b\,S_{b\cdt}^{\fdt c}\equiv0.$ It is
seen that RHS of (\ref{spin}) can be represented as
$\hbar/2\bps\gam^3\psi$ which is eigenvalue of operator of
helicity $\hbar/2\gam^3$  in state described by wave function
$\psi$ in the tangent polarization space. Moreover, it can be
shown that definition (\ref{spin}) is in accord with formula for
"$-$"-component of current of spin derived from Noether theorem in
field theory \cite{NN}.

Due to the definition of the generalized momentum $p_a$ given
above the one can be represented as follows:
\begin{equation}\label{p_a}
p_a=E\,\bps\psi\,\eta_{ab}\dtx^b+\hspin\gamma_{a\delta\eps}S^{\delta\eps}=
\pi_a+\hspin\gamma_{a\delta\eps}S^{\delta\eps}.
\end{equation}
It is seen that form of (\ref{p_a}) coincides with analogous
equation obtained for massive particle of spin $1/2$ \cite{al}.
\comment{All differences between the equations are values of
indexes of spin which now run over numbers $\{1,2\}$ and energy
$E$ included to $\pi_a.$ instead mass $m$.}

According to the equation (\ref{eq4psi}) straightforward
calculation of covariant derivative of spin (\ref{spin}) gives:
\begin{equation}\label{DS:ds}
\Ds{S^{\delta\eps}}=\dds{S^{\delta\eps}}
+\dtx^a\lrr{\gamma_{a\gamma\cdt}^{\fdt\fdt \delta}S^{\gamma\eps}
+\gamma_{a\gamma\cdt}^{\fdt\fdt \eps}S^{\delta\gamma}}\equiv0,
\end{equation}
where we took into account the fact that total covariant
derivatives of Pauli matrices are zero.

\section{Euler-Lagrange equations for $x^i$ variables}

Euler-Lagrange equations for $x^i$ variables read:
\begin{equation}\label{ELEQ}
{dp_i\,/ds}=\pp{\Lge}{x^i}.
\end{equation}
Under comparison (\ref{lagr}) with the Lagrangian of spin $1/2$
massive particle \cite{al} we see that parts of the Lagrangians
contributing to $\pp{\Lge}{x^i}$ formally coincide. At the same
time, as it was pointed above expressions for the generalized
momenta both massless and massive particles of spin $1/2$ are
formally identical. Noting that (\ref{DS:ds}) has the same form as
equation for spin of massive particle \cite{al} we expect that
Euler-Lagrange equations (\ref{ELEQ}) will be reduced to the same
form as in case of massive particle \cite{zr1,al}:
$$D\pi_a/ds=\hsp R_{\delta\eps ab}\,\dtx^bS^{\delta\eps},
$$ where
$$\cur{c}{d}=d\con{c}{d}+\con{e}{d}\wdg\con{c}{e}=1/2R_{c\cdt
ab}^{\fdt d}\,\nu^a\wdg\nu^b
$$ is 2-form of curvature.
Now substituting $\pi_a=E\bps\psi\,\eta_{ab}\dtx^b$ into the above
equation and keeping in mind fact that the coefficient at
$\eta_{ab}\dtx^b$ is constant we rewrite the equation as follows:
\begin{equation}\label{papa}
E\bps\psi\,\Ds{\dtx^a}=\hsp R^{\fdt\fdt a}_{\delta\eps\cdt
b}\,\dtx^bS^{\delta\eps}.
\end{equation}
It is seen that (\ref{DS:ds}) and (\ref{papa}) constitute set of
equations of motion of massless particle of spin $s=1/2$ which is
similar to system of Papapetrou equations for motion of photon in
curved space-time as they presented in the work \cite{za}.

\section*{Acknowledgment}
The author express his gratitude to  Z. Ya. Turakulov who
motivated the author to carry out these studies and whose critical
remarks provided significant improve of present article. This
research was supported by project FA-F2-F061 of Uzbekistan Academy
of Sciences.
%the grant AC-83 of the Abdus~Salam International Center for
%Theoretical Physics.

%\section*{References}
\end{document}